\documentclass[tighten,twocolumn,useAMS,appendixfloats,apj]{likeapj}
\usepackage{etoolbox}
\makeatletter
\patchcmd{\NAT@citex}
  {\@citea\NAT@hyper@{
     \NAT@nmfmt{\NAT@nm}
     \hyper@natlinkbreak{\NAT@aysep\NAT@spacechar}{\@citeb\@extra@b@citeb}
     \NAT@date}}
  {\@citea\NAT@nmfmt{\NAT@nm}
   \NAT@aysep\NAT@spacechar\NAT@hyper@{\NAT@date}}{}{}

\patchcmd{\NAT@citex}
  {\@citea\NAT@hyper@{
     \NAT@nmfmt{\NAT@nm}
     \hyper@natlinkbreak{\NAT@spacechar\NAT@@open\if*#1*\else#1\NAT@spacechar\fi}
       {\@citeb\@extra@b@citeb}
     \NAT@date}}
  {\@citea\NAT@nmfmt{\NAT@nm}
   \NAT@spacechar\NAT@@open\if*#1*\else#1\NAT@spacechar\fi\NAT@hyper@{\NAT@date}}
  {}{}

  \makeatother
\usepackage{mathptmx}

\received{\today}
\revised{}
\accepted{}
\published{}
\submitjournal{\apj}

\shorttitle{}
\shortauthors{R. Serafinelli et al.}

\usepackage{graphicx,natbib,bm,times,amsmath,float}

\hypersetup{linkcolor=blue,citecolor=blue,filecolor=cyan,urlcolor=blue,bookmarksnumbered=true,bookmarksopen=true}
\usepackage[all]{hypcap}

\include{newcomm}   

\begin{document}
\title{Time-evolving diagnostic of the ionized absorbers in NGC 4051\\
I. High-resolution time-averaged spectroscopy}
\author{Roberto Serafinelli}
\affil{Instituto de Estudios Astrofísicos, Facultad de Ingeniería y Ciencias, Universidad Diego Portales, Avenida Ejército Libertador 441, Santiago, Chile}\footnote{\email{roberto.serafinelli@mail.udp.cl}}
\affil{INAF - Osservatorio Astronomico di Roma, Via Frascati 33, 00078, Monte Porzio Catone (Roma), Italy}
\author{Fabrizio Nicastro}
\affil{INAF - Osservatorio Astronomico di Roma, Via Frascati 33, 00078, Monte Porzio Catone (Roma), Italy}
\author{Alfredo Luminari}
\affil{INAF - Istituto di Astrofisica e Planetologia Spaziali, Via del Fosso del Cavaliere 100, 00133, Roma, Italy}
\affil{INAF - Osservatorio Astronomico di Roma, Via Frascati 33, 00078, Monte Porzio Catone (Roma), Italy}
\author{Yair Krongold}
\affil{Instituto de Astronomía, Universidad Nacional Autónoma de México, Circuito Exterior, Ciudad Universitaria, Ciudad de México 04510, México}
\author{Francesco Camilloni}
\affil{INAF - Istituto di Astrofisica e Planetologia Spaziali, Via del Fosso del Cavaliere 100, 00133, Roma, Italy}
\author{Elias Kammoun}
\affil{Cahill Center for Astrophysics, California Institute of Technology, 1216 East California Boulevard, Pasadena, CA 91125, USA}
\author{Riccardo Middei}
\affil{Center for Astrophysics - Harvard \& Smithsonian, 60 Garden Street, Cambridge, MA, 02138, USA}
\affil{INAF - Osservatorio Astronomico di Roma, Via Frascati 33, 00078, Monte Porzio Catone (Roma), Italy}
\affil{Space Science Data Center, Agenzia Spaziale Italiana, Via del Politecnico snc, 00133, Roma, Italy}
\author{Enrico Piconcelli}
\affil{INAF - Osservatorio Astronomico di Roma, Via Frascati 33, 00078, Monte Porzio Catone (Roma), Italy}
\author{Luigi Piro}
\affil{INAF - Istituto di Astrofisica e Planetologia Spaziali, Via del Fosso del Cavaliere 100, 00133, Roma, Italy}

\begin{abstract}
  We present a high-resolution X-ray spectroscopic study of the Narrow-Line Seyfert 1 galaxy NGC 4051 using two \textit{XMM-Newton} high-resolution Reflection Grating Spectrometer (RGS) observations. The spectra reveal three distinct layers of photoionized gas flowing outward from the central black hole: a low-ionization phase (LIP), a higher-ionization phase (HIP), and a high-velocity and high ionization phase (HVIP). Each absorber leaves characteristic imprints on the soft X-ray spectrum. While the LIP and HVIP are fully consistent with being in ionization equilibrium with the central radiation field over the course of the $\sim$250 ks spanned by the two observations, the HIP shows a significant change in ionization ($3.8\sigma$), suggesting non-equilibrium. By modeling the two spectra with our time-dependent photoionization code ({\footnotesize TEPID}), we constrain the density of the HIP gas to $\log n_{\rm H}=7.7^{+0.2}_{-0.9}$ and estimate its distance to be about $R=0.45^{+0.80}_{-0.09}$ light-days from the black hole, corresponding to $R=4000^{+7000}_{-800}$ gravitational radii. In contrast, the narrow soft X-ray emission lines remain constant, consistent with an origin in the more extended narrow-line region. Our results show the value of combining high-resolution and time-resolved spectroscopy to probe the structure, physical conditions, and variability of AGN outflows.

\end{abstract}

\keywords{ galaxies: active -- galaxies: individual: NGC 4051 – X-rays: galaxies.}

\section{Introduction}
\label{sec:intro}

Active Galactic Nuclei (AGN) are among the most luminous and energetic objects in the universe, fueled by the accretion of matter onto supermassive black holes at their centers. The vast amounts of gravitational energy released in this process produce intense radiation across the electromagnetic spectrum, with X-rays providing a key observational signature. The primary X-ray emission in AGN is typically well described by a cut-off power law \citep[e.g.,][]{fabian15,tortosa18,kamraj22,serafinelli24}, which originates from inverse Compton scattering of soft UV photons by energetic electrons in a hot corona with temperatures reaching \(T\sim10^9\) K \citep[e.g.,][]{haardt91,haardt93}.\\
\indent The intrisic X-ray continuum in AGN provides a direct probe of the physical conditions near the accreting black hole. This emission is often modified by intervening material along the line of sight. One of the most prominent absorption components observed in AGN spectra is the so-called Warm Absorber (WA), caused by partially ionized gas that imprints distinctive absorption features in the soft X-ray band \citep[e.g.,][]{halpern84}. WAs are detected in approximately \(50-60\%\) of local type 1 AGN \citep[e.g.,][]{blustin05,piconcelli05,mckernan07,winter12,laha14,laha16}, appearing as outflows with moderate ionization states, column densities of the order of \(N_{\rm H}\sim10^{21}\) cm\(^{-2}\), and velocities typically ranging between \(v\sim500-1000\) km s\(^{-1}\).\\ 
\indent Standard photoionization equilibrium models assume that the so-called "equilibration timescale" $t_{eq}$ \citep{nicastro99} with which the gas readjusts to luminosity variations is significantly shorter than the variability timescale of the ionizing source. Such timescale is inversely proportional to the electron number density $n_e$. When $n_e$ is low enough that $t_{eq}$ become comparable to the variability timescale, the equilibrium assumption is no longer appropriate and it is thus necessary to properly follow the out-of-equilibrium ionization evolution of the gas \citep{krolik95,nicastro99,krongold07,sadaula23}. Recent advancements in time-evolving photoionization codes, such as {\footnotesize TPHO} \citep{rogantini22}, {\footnotesize TEPID} \citep{luminari23}, and updated versions of {\footnotesize XSTAR} \citep{sadaula23}, have enabled detailed, self-consistent treatment of these non-equilibrium conditions, by following the temporal evolution of ionic concentrations and electron temperature as a function of the incident (time-variable) ionizing lightcurve. By comparing the resulting absorption spectrum with time-resolved observed spectra, it is then possible to directly constrain $n_e$ by looking at the temporal evolution of the gas spectral features, chiefly the overall amplitude of the variations and their temporal delay with respect to the primary variability \citep{luminari23,kosec24}. Once $n_e$ and the ionization parameter $U={Q_{\rm ion}}/{4\pi c R^2 n_{\rm H}}$, where $Q_{\rm ion}$ is the ionizing luminosity (in units of photons) and $n_{\rm H}$ is the hydrogen-equivalent number density ($n_{\rm H}\approx1.2 n_e$ for standard solar abundances), are constrained, it is then possible to derive the distance of the gas from the ionizing source $R$ as $R= \sqrt{{Q_{\rm ion}}/{4\pi c U n_{\rm H}}}$. In the standard equilibrium analysis, however, $n_{\rm H}$ is a degenerate quantity and, thus, $R^2$ and $n_{\rm H}$ cannot be disentangled. In that case, lower limits to the density, hence upper limits on the distance, are typically assumed. A complementary approach to constrain the density, and therefore break the $n_{\rm H}$ - $R$ degeneracy, involves the detection of absorption lines arising from metastable levels in ions such as O~{\footnotesize V} or Fe~{\footnotesize XIX}, whose population depends on the local density \citep[e.g.,][]{steinbrugge22}. However, such transitions have not yet been unambiguously observed in AGN X-ray spectra, possibly due to densities being below the critical thresholds needed to populate these levels \citep[e.g.,][]{kaastra04}.\\
\indent The Narrow-line Seyfert 1 galaxy NGC 4051 ($z=0.00234$), has been a pivotal target in the study of outflows, due to the presence of winds in the UV \citep[e.g.,][]{kraemer12} and X-rays \citep[e.g.,][]{pounds04,krongold07,steenbrugge09,lobban11}. The source also hosts a small-scale jet \citep[e.g.,][]{berton18,jarvela22}. 
Due to its large and fast soft X-ray variable luminosity, ascribed to its relatively low black hole mass \citep[$M_{\rm BH}\sim1.4\times10^6$ $M_\odot$][]{peterson00}, NGC 4051 is well suited for temporal-evolution studies of its multiple warm absorbers. High-resolution spectroscopy with \textit{XMM-Newton} Reflection Gratings Spectrometer (RGS) and \textit{Chandra} Transmission Gratings has indeed revealed the presence of multiple WA components. These differ by nearly two orders of magnitude in ionization parameter, some of which exhibiting significant temporal variations, likely ascribed to the variation of the ionization state of the gas \citep{pounds04,krongold07,steenbrugge09}.\\ 
\indent \cite{krongold07} (hereafter K07) used time-averaged RGS spectroscopy to identify two WA components with consistent outflow velocity of $\sim 500$ km s$^{-1}$ and different ionization states, a low- and a high-ionization phases, (LIP and HIP, respectively). They also performed a time-dependent analysis of the response of these absorbers to the ionizing continuum to constrain the electron density of the gas. This placed the HIP at a distance of $0.5–1.0$ light-days ($\sim 1100-2200$ $R_{\rm g}$, where $R_{\rm g}=GM/c^2$ is the gravitational radius of the source) and the LIP within $3.5$ light-days ($<7900$ $R_{\rm g}$) from the ionizing source. This strongly suggests an origin in an accretion disk wind rather than in the dusty torus or the narrow-line region, located not closer than 12 light-days.\\ 
\indent \cite{steenbrugge09} report also multiple ionized absorbers in NGC 4051 with the {\it Chandra} LETG. In particular, the authors detected the presence of a fast absorber with $v_{\rm out}=4670\pm150$ km s$^{-1}$ and ionization $\log\xi/({\rm erg\; cm\; s}^{-1})=3.1\pm0.1$, corresponding to $\log U\simeq1.6$ (assuming the spectral energy distribution (SED) from \cite{crenshaw12}). 
The same absorber was also detected by \cite{lobban11} with $v_{\rm out}=5800^{+860}_{-1200}$ km s$^{-1}$ and $\log\xi=4.1^{+0.2}_{-0.1}$, corresponding to $\log U\simeq2.6$.\\
\indent In this paper and in a forthcoming one (Luminari et al., in prep. - Paper II) we present a detailed analysis of two {\it XMM-Newton} observations and a long {\it NuSTAR} observation embedding the \textit{XMM-Newton} ones. In Paper I we present the high-resolution analysis of the RGS data from the two observations, while Paper II will present a time-resolved EPIC-pn and {\it NuSTAR} spectral analysis of the ionized absorbers identified in the present paper. Time-averaged low-resolution spectroscopy of the same observations presented here, aimed at the study of an X-ray flare in the light curve (see Fig.~\ref{fig:lc}), was presented in \cite{kumari23}. Sect.~\ref{sec:data} describes the data and their reduction process in detail. Sect.~\ref{sec:rgs} presents the spectroscopy of the RGS camera. In Sect.~\ref{sec:disco} we discuss the results of our analysis, presenting estimates for the location and density of the absorbers. In Sect.~\ref{sec:conclusions} we summarize our results. 
Throughout the paper, we adopt a standard flat $\Lambda$CDM cosmology with $h_0=70$ km s$^{-1}$ Mpc$^{-1}$. $\Omega_m=0.3$ and $\Omega_\Lambda=0.7$.

\section{Data}  
\label{sec:data}
    
\begin{table}
\centering
\caption{Journal of the X-ray observations analyzed in this work. Since RGS 1 and RGS2 were combined in the analysis we only report the sum of the exposure of the two detectors. Instead, for FPMA and FPMB we reported the exposures separately.}
\label{tab:data}
\begin{tabular}{lcccccc}
\hline \hline
Epoch & OBSID & Instrument & Date & Net exp. (s)\\
\hline
All & 60401009002 & FPMA & 2018-11-04 & 311139\\
& & FPMB & & \\
1 & 0830430201 & {EPIC-pn} & 2018-11-07 & 48750\\
& & RGS1+2 & & 160400\\
2 & 0830430801 & {EPIC-pn} & 2018-11-09 & 51420\\
& & RGS1+2 & & 168300\\
\hline
\end{tabular}
\end{table}

\begin{figure}
\centering
\includegraphics[scale=0.37]{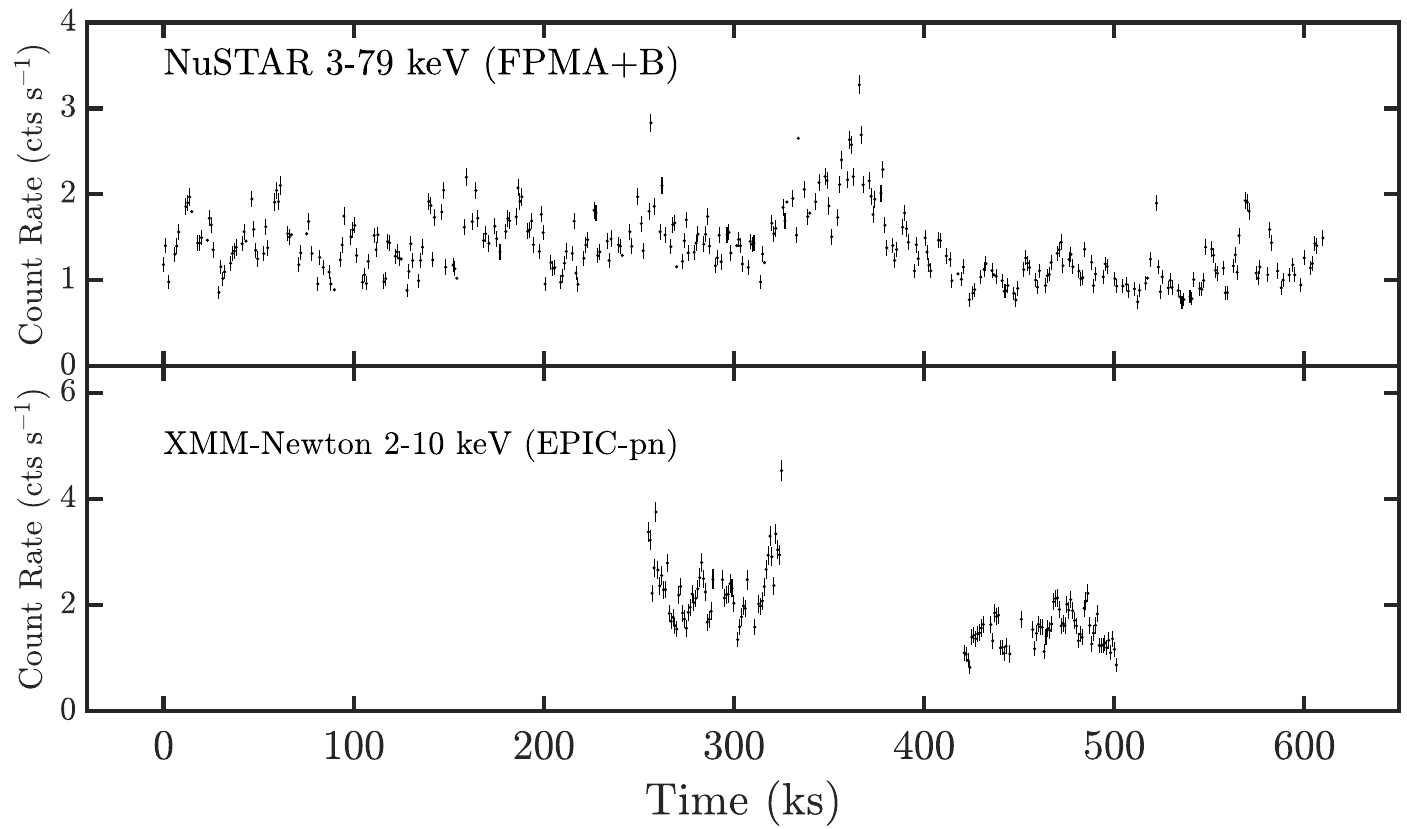}
\caption{{\it Upper panel}. {\it NuSTAR} combined FPMA+B background-subtracted light curve in the $3-79$ keV energy band. {\it Lower panel}. EPIC-pn light curves of both observations in the $2-10$ keV energy band. All light curves are binned at $1$ ks per time bin.}
\label{fig:lc}
\end{figure}

NGC~4051 was observed by {\it XMM-Newton} between November 7--9 2018, with two $\sim80$~ks pointings (Epochs 1 and 2, hereinafter), both overlapping with a longer (4--10 November 2018: $\sim310$~ks net exposure) {\it NuSTAR} observation (Table~\ref{tab:data} and Fig.~\ref{fig:lc}). 
{\it NuSTAR} data were reduced with the standard {\footnotesize NUPRODUCTS} task of the {\footnotesize NUPIPELINE} software package included in {\footnotesize HEASOFT} v6.30. 
We used the calibration files available in the {\footnotesize CALDB} repository up to version $20240311$. The two Focal Plane Modules A and B (FPMA/B) spectra were extracted with the task {\footnotesize NUPRODUCTS} from a circular region with radius $60"$ centered on the source, with backgrounds extracted from two source-free circular regions of $40"$ radius for each module. Light curves for the combined FPMA+FPMB modules in the $3-79$ keV energy band were extracted and are shown in the upper panel of Fig.~\ref{fig:lc} with $1$-ks binning. While we will need such light curve for the TEPID simulations, the spectra will instead be analyzed together with the EPIC-pn in Paper II.\\
\indent  All {\it XMM-Newton} data-products were extracted with {\footnotesize SASv20.0.0}. For the RGS data, whose analysis is presented here, we used the standard processing tool {\footnotesize RGSPROC}, while the response files were generated with {\footnotesize RGSRMFGEN}. To increase the signal-to-noise per resolution element, data from the two RGS cameras were combined in a single spectrum  using {\footnotesize RGSCOMBINE}. We adopt the native RGS binning scheme ($\Delta\lambda = 0.01$ \AA\, i.e. $\sim 6$ bins per resolution element) for the fitting. \\
\indent EPIC-pn data were processed with the standard tool {\footnotesize EPPROC}. We create good time intervals to veto proton flaring events using {\footnotesize TABGTIGEN}, after which net exposures are reduced to $\sim48$ ks for Epoch 1 and to $\sim51$ ks for Epoch 2. Pile-up is negligible, due to the observation being performed in small window. We extracted EPIC-PN light curves from a region with 40" radius, centered around the source, while the background was extracted from a source-free region with the same size. The light curves of the two observations in the $E=2-10$ keV energy range are also shown in Fig.~\ref{fig:lc}. We also retrieved data from the XMM-{\it Newton} Optical Monitor \citep{mason01}. During the $2018$ observations, NGC 4051 was observed using only the U filter ($3500$ $\AA$) throughout the monitoring period. The OM data were processed following the standard approach in SAS using the task {\footnotesize OMCHAIN}, and the spectral data were converted into an {\footnotesize XSPEC} \citep{arnaud96} compatible format with the task {\footnotesize OM2PHA}.

\section{RGS spectral analysis}
\label{sec:rgs}

\begin{table}[]
    \centering
    \caption{Continuum values before the addition of emission and absorption components}
    \label{tab:continuum}
    \begin{tabular}{lcc}
    \hline \hline
    Parameter & Epoch 1 & Epoch 2\\
    \hline
    {\bf Black Body}\\
     $kT$ (eV) & $101^{+2}_{-1}$   & $-$ \\
     norm ($10^{-4}$ cts s$^{-1}$ cm$^{-2}$ keV$^{-1}$) & $1.2\pm0.1$ & $1.1\pm0.1$\\ 
     {\bf Power Law}\\
     $\Gamma$ & $1.95\pm0.07$ & $1.64\pm0.11$\\
     norm ($10^{-3}$ cts s$^{-1}$ cm$^{-2}$ keV$^{-1}$) & $5.5\pm0.1$ & $1.8\pm0.1$\\
     $C/{\rm dof}$ & 7227/6276\\
     \hline
    \end{tabular}
\end{table}

\begin{table}
    \centering
    \caption{Best-fit parameters of the RGS spectra, assuming a model with three ionized absorbers: the high- and low- ionization phases (HIP and LIP, respectively), and the high-velocity and ionization phase (HVIP). Values omitted for Epoch 2 imply that the value is kept constant with Epoch 1. The column density of the ionized absorbers is assumed constant, only allowing the ionization parameter to vary. Concerning the emission lines, the centroid wavelength and normalizations are kept tied between observations, while the widths are unresolved in RGS ($\sigma=0$).}
    \begin{tabular}{lcccccccc}
    \hline \hline
     Parameter    &  Epoch 1 & Epoch 2\\
     \hline 
     {\bf HIP}\\
     $\log U$ & $1.23^{+0.07}_{-0.05}$ & $1.12\pm0.04$\\
     $N_{\rm H}$ ($10^{21}$ cm$^{-2}$) & $1.4^{+0.2}_{-0.3}$ & $-$\\
     $v_{\rm out}$ (km s$^{-1}$) & $530\pm70$ & $-$\\
      {\bf LIP}\\
     $\log U$ & $-1.0\pm0.3$ & $-1.1^{+0.6}_{-0.4}$\\
     $N_{\rm H}$ ($10^{21}$ cm$^{-2}$) & $0.16^{+0.05}_{-0.06}$ & $-$\\
     $v_{\rm out}$ (km s$^{-1}$) & $400\pm120$ & $-$\\
     {\bf HVIP}    &    \\
     $\log U$ & $2.3^{+0.1}_{-0.2}$ & $2.1^{+0.2}_{-0.3}$\\
     $N_{\rm H}$ ($10^{21}$ cm$^{-2}$) & $2.7^{+1.4}_{-1.6}$ & $-$\\
     $v_{\rm out}$ (km s$^{-1}$) & $5800\pm300$ & $-$\\
     {\bf Black Body}\\
     $kT$ (eV) & $101\pm3$ & $-$\\
     norm ($10^{-5}$ cts s$^{-1}$ cm$^{-2}$ keV$^{-1}$) & $6\pm2$ & $5\pm1$\\
     {\bf Power Law}\\
     $\Gamma$ & $2.24^{+0.07}_{-0.06}$ & $2.09^{+0.12}_{-0.14}$\\
     norm ($10^{-3}$ cts s$^{-1}$ cm$^{-2}$ keV$^{-1}$) & $6.7^{+0.2}_{-0.3}$ & $3.8\pm0.2$\\
     {\bf O {\footnotesize VII} Emission Triplet}\\
     Centroid Wavelength 1 ($\AA$) & $22.114^{+0.005}_{-0.008}$ & $-$\\
     norm 1 ($10^{-5}$ cts s$^{-1}$ cm$^{-2}$ keV$^{-1}$) & $7\pm1$ & $-$\\
     Centroid Wavelength 2 ($\AA$) & $21.74^{+0.01}_{-0.03}$ & $-$ \\
     norm 2 ($10^{-5}$ cts s$^{-1}$ cm$^{-2}$ keV$^{-1}$) & $1.6^{+0.9}_{-0.8}$ & $-$\\
     Centroid Wavelength 3 ($\AA$) & $21.60^{+0.06}_{-0.05}$ & $-$ \\
     norm 3 ($10^{-5}$ cts s$^{-1}$ cm$^{-2}$ keV$^{-1}$) & $2.3^{+0.9}_{-0.8}$ & $-$\\
     {\bf O {\footnotesize VIII} K$\alpha$ Emission Line} \\
     Centroid Wavelength ($\AA$) & $18.96^{+0.01}_{-0.03}$ & $-$\\
     norm ($10^{-5}$ cts s$^{-1}$ cm$^{-2}$ keV$^{-1}$) & $2.5\pm0.5$ & $-$\\
     {\bf Ne {\footnotesize IX} Emission Line}\\
     Centroid Wavelength ($\AA$) &  $13.675^{+0.005}_{-0.023}$ & $-$\\
     norm ($10^{-5}$ cts s$^{-1}$ cm$^{-2}$ keV$^{-1}$) & $1.0\pm0.5$ & $-$ \\
     C/dof & $6648/6254$ & \\
     \hline
    \end{tabular}
    \label{tab:rgs_2abs}
\end{table}

\begin{figure*}
\centering 
\includegraphics[scale=0.7]{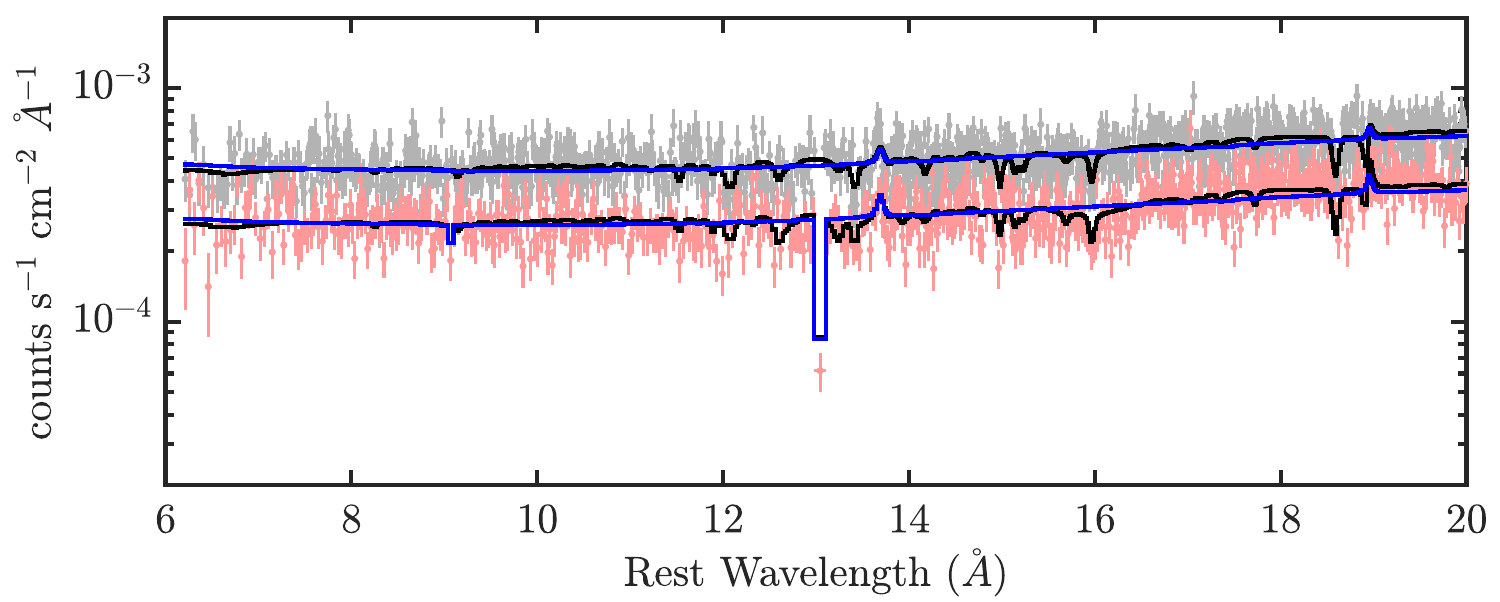}
\includegraphics[scale=0.7]{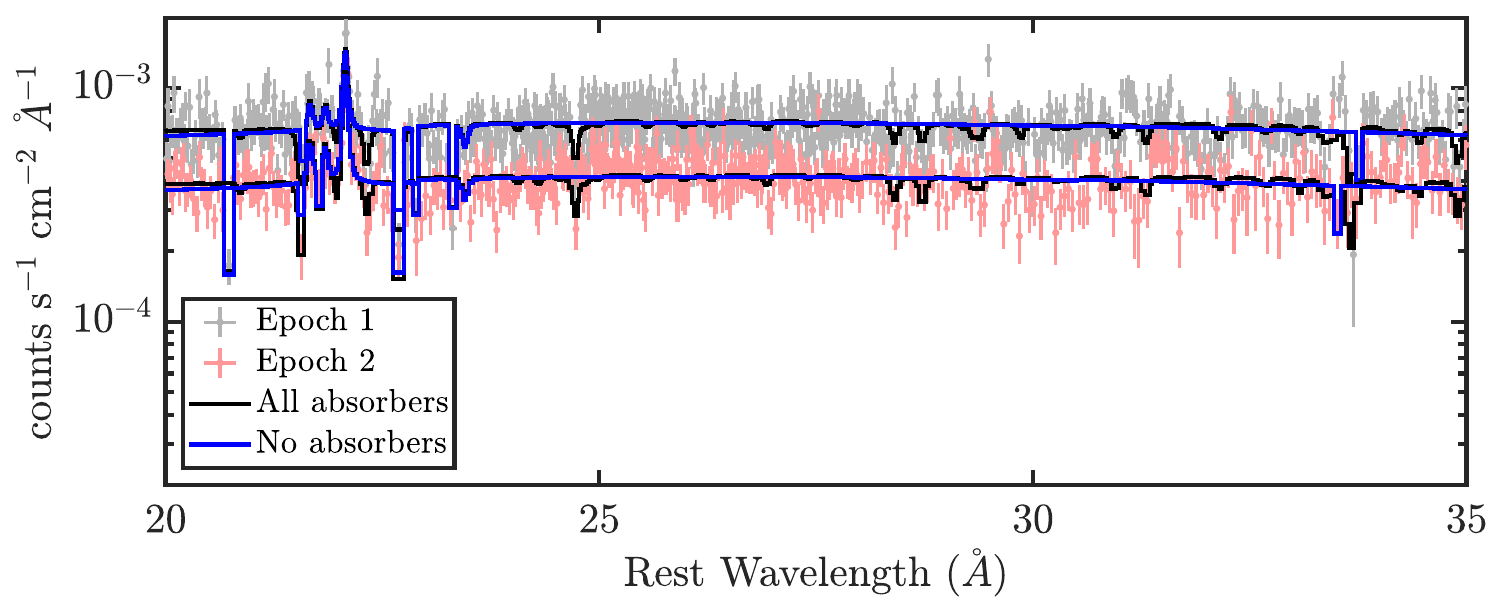}
\caption{The $\lambda=6-35$ \AA $\,$ RGS spectra analyzed in this paper. The grey spectrum is Epoch 1, while the light red spectrum is the one taken in Epoch 2. The black line is the best-fit model that includes all the absorbers detected in this paper. We also include the model without the absorbers, in blue, that also includes Gaussian emission lines, showing that there are significant residuals that strongly point at the presence of several ionized absorbers. The sharp vertical dips are instrumental gaps. For visual purposes only, the spectral data points have been rebinned at a minimum of 5 counts per wavelength bin.}
\label{fig:rgs_obs1}
\end{figure*}

Here we analyze the RGS spectra of both Epochs 1 and 2. In the $\Delta\lambda = 10$ m\AA\, native-binning scheme adopted here, not all spectral bins have a sufficiently high number of counts for the use of the $\chi^2$ statistic \citep[e.g.,][]{kaastra17}, and we therefore adopt the Cash statistic \citep{cash79}. We consider the range $\lambda=6-35\ \AA$ ($E=0.35-2.1$ keV) in the spectral fitting.\\
\indent We first fit the spectra with a simple power-law plus black-body continuum model, absorbed by the Galactic cold gas (N$_H = 1.19 \times 10^{20}$ cm$^{-2}$, \citealp{hi4pi16}). The black body models the soft excess observed in the source, and its temperature is kept constant during the campaign, as done in previous analysis, \citet{pounds04,uttley04}, and as confirmed in Paper II, while the power-law spectral index and the normalizations of the two components are left free to vary independently. The best-fitting parameters are listed in Table~\ref{tab:continuum}. This model resulted in a statistic of $C/{\rm dof}=7227/6276$, where dof are the degrees of freedom. Several line-like absorption and emission features are clearly left unfitted in the data by a pure-continuum model (see Fig.~\ref{fig:rgs_obs1}). In particular, in both epochs, positive residuals are seen at $\lambda\sim 21.6-22.1 \AA$ and $\lambda\sim13.5\AA$, suggesting the presence of O {\footnotesize VII} and Ne {\footnotesize IX} emission. In Epoch 2 (lower continuum flux), positive residuals also emerge at $\sim19$ \AA, indicating the presence O {\footnotesize VIII} emission. We include five emission Gaussians to model these narrow positive features: three for the putative O {\footnotesize VII} triplet, one for the strongest line of the Ne {\footnotesize IX} triplet and one for the unresolved O {\footnotesize VIII} doublet. Centroids and normalizations of the Gaussians are left free to vary, while their widths are all frozen at the RGS resolution $\sigma=10$ m\AA. Finally, Gaussian centroids were kept tied between the two epochs. The best-fitting emission-line normalizations are found to be fully consistent between the two epochs, suggesting no variation of the line emissivities over the explored timescale of $\sim 250$ ks, as often seen in AGN over such timescales \citep[e.g., ][]{bianchi06,bianchi19,grafton-waters21}. We therefore tie the line normalization to common values between the two epochs, and refit the data, obtaining the best-fitting parameters listed in Table~\ref{tab:rgs_2abs}. The three emission lines that make up the positive residuals at $\sim21-22$ \AA\ are found at $\lambda=22.114^{+0.005}_{-0.008}$ \AA, $21.74^{+0.01}_{-0.03}$ \AA, $21.60^{+0.06}_{-0.05}$\AA, fully consistent with an O {\footnotesize VII} K$\alpha$ complex, precisely with the forbidden, intercombination and recombination lines of the triplet, respectively. A fourth line cures the $\sim19$ \AA\ residuals in Epoch 2 and is found at $\lambda=18.96^{+0.01}_{-0.03}$ \AA, consistent with O {\footnotesize VIII} K$\alpha$ emission. The fifth emission line is found at $\lambda=13.675^{+0.005}_{-0.023}$ \AA, and is identified with the forbidden Ne {\footnotesize IX} K$\alpha$ emission. 
After the inclusion of these five emission lines, the statistic improves to $C/{\rm dof}=6964/6266$ ($\Delta C/\Delta {\rm dof}=263/10$).\\ 
\indent Several negative residuals are also present. We model these features by adding one photoionized absorber via the {\footnotesize PHASE} table model \citep{krongold03}, consisting in a spectral interface routine that computes absorption spectra, made of Voight absorption lines and continuum opacity, as a function of tabulated ionic abundances. Such abundances are pre-computed with {\footnotesize CLOUDY} \citep{ferland98} as a function of the gas ionization $U$ and the (hydrogen-equivalent) column density $N_{\rm H}$. We simulated ionizations from $\log U=-2$ to $\log U=4$ with steps of $0.05$), and $\log(N_{\rm H}/cm^{-2})$ from $19$ to $24$ with steps of 0.5. We adopt the observed UV-to-X-ray spectral energy distribution (SED), as derived from the {\it XMM-Newton} (OM, RGS and EPIC-pn) and {\it NuSTAR} data. As the source is variable in the X-rays, we assume the average flux between the two observations. {\footnotesize PHASE} has then four free parameters: $\log U$, $\log N_{\rm H}$, the gas turbulent velocity $v_{\rm turb}$  and the observer-frame redshift $z_{\rm obs}$. We keep $N_{\rm H}$ tied between the two Epochs, as it is quite unlikely for the column density to vary in such a short timescale, and we leave $\log U$ free to vary. We assume $v_{\rm turb}=200$ km s$^{-1}$, unresolved at RGS resolution and we tie the redshift\footnote{The absorber blue/red-shift in the source rest frame is $z_a=(1+z_{\rm obs})/(1+z_c)-1$, where $z_c$ is the cosmological redshift. The outflow (or inflow) velocity is then given by $v_{\rm out}/c=(z_a^2+2z_a)/(z_a^2+2z_a+2)$.} between the two Epochs. The statistic improves to $C/{\rm dof}=6752/6262$ ($\Delta C/\Delta {\rm dof}=212/4$). The column density of this absorber is given by $N_{\rm H}=1.4^{+0.2}_{-0.3}\times10^{21}$ cm$^{-2}$. The best-fitting ionization parameters are $\log U=1.23^{+0.07}_{-0.05}$ for Epoch 1 and $\log U=1.12\pm0.04$ for Epoch 2, i.e. consistent with each other at $90\%$ confidence level. The velocity of the absorber is $v_{\rm out}=530\pm70$ km s$^{-1}$, consistent with the values found by K07 for both their components. Given the high value of $U$, we dub this absorber High-Ionization Phase (HIP), as in K07.\\
\indent A visual inspection of the residuals shows that several absorption lines are still not accurately modeled by the single {\footnotesize PHASE} screen (see Fig.~\ref{fig:rgs_obs1}), suggesting the presence of a second absorber with different ionization. Indeed, the addition of a second {\footnotesize PHASE} component to our model, with the same parameter-constraints described for the HIP, further improves the statistic to $C/{\rm dof}=6697/6258$ ($\Delta C/\Delta {\rm dof}=55/4$). The best-fitting outflow velocity of this absorber is consistent with that of the HIP: $v_{\rm out}=400\pm120$ km s$^{-1}$. Its best-fitting column density is $N_{\rm H}=(1.6^{+0.5}_{-0.6})\times10^{20}$ cm$^{-2}$, while the ionization parameters are $\log U=-1.0 \pm0.3$ and $\log U=-1.1^{+0.6}_{-0.4}$ in Epoch 1 and 2, respectively, again fully consistent with each other within the uncertainties. Given the lower value of $U$ with respect to the HIP and the consistent value of $v_{out}$ we identify this component with the Low-Ionization Phase (LIP) in K07.\\
\indent A few additional residuals are still present in the spectra after the addition of the LIP absorber, particularly around the Fe L and O {\footnotesize VII-VIII} K transitions. We therefore add a third PHASE component, with the same parameter-constraints of LIP and HIP. The statistic is $C/{\rm dof}=6648/6254$, implying an improvement of $\Delta C/\Delta{\rm dof}=49/4$. $v_{\rm out}=5800\pm300$ km s$^{-1}$ is an order of magnitude larger than that of LIP and HIP and consistent with one of the components reported by \cite{steenbrugge09} and \cite{lobban11} (see Sect.~\ref{sec:hvip_comparison} for a detailed comparison). Its column density is $N_{\rm H}=2.7^{+1.4}_{-1.6}\times10^{21}$ cm$^{-2}$, while the ionization parameters are $\log U=2.3^{+0.1}_{-0.2}$ and $\log U=2.1^{+0.2}_{-0.3}$ in Epoch 1 and Epoch 2, respectively. Given its large velocity and ionization, we dub this absorber High-Velocity and -Ionization Phase (HVIP).\\
\indent The addition of the photoionized absorbers does not affect the best-fit values of the black body temperature nor the emission lines, but affects the power law photon index, which is now $\Gamma=2.24^{+0.07}_{-0.06}$ in Epoch 1, and $\Gamma=2.09^{+0.12}_{-0.14}$ in Epoch 2, significantly different from the continuum-only values reported in Table~\ref{tab:continuum}. The best fit values of this model are listed in Table~\ref{tab:rgs_2abs}, while the best-fitting model is shown in Fig.~\ref{fig:rgs_obs1}.\\

\section{Discussion}
\label{sec:disco}

\subsection{Contribution of the Ionized Absorbers to the RGS Spectrum}
\label{sec:absorbers}

\begin{figure*}
    \centering
    \includegraphics[scale=0.7]{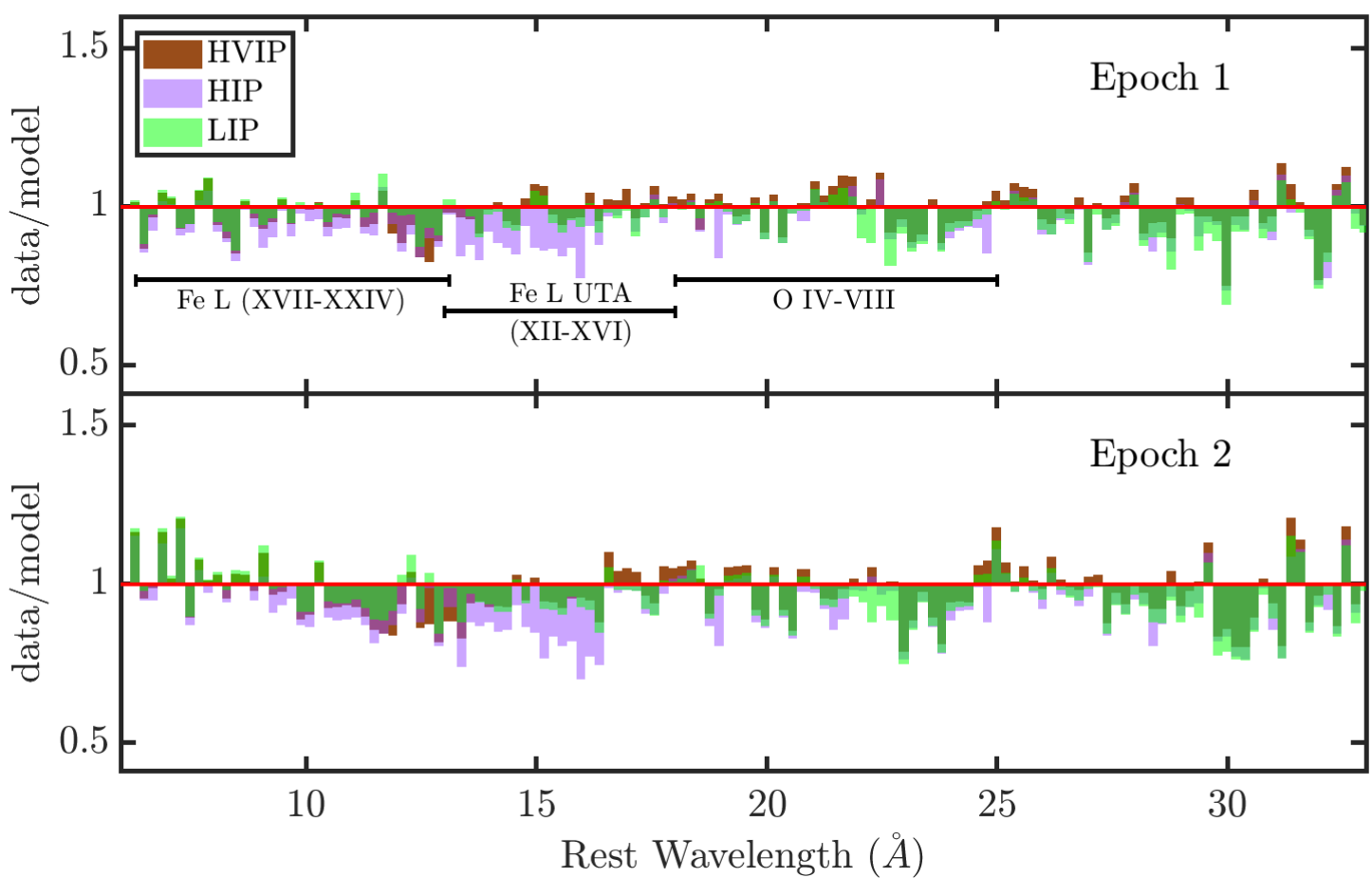}
    \caption{Data-to-model ratios that highlight the contributions to the fit from the three ionized absorbers. The horizontal line in red represents the best-fit model described in Table~\ref{tab:rgs_2abs}. We removed one absorber at a time to highlight the difference between the best-fit model and the model that does not include the absorber. The green shaded area represents residuals obtained by removing the LIP, the lavender area is obtained when we remove the HIP, while the brown area is obtained by removing the HVIP. The upper panel represents Epoch 1 and the lower panel represents Epoch 2. We stress that the data are not refit after removing each absorber and are just plotted for illustrative purposes. The data is also visually binned at a minimum of 20 counts per wavelength bin.}
    \label{fig:contrib_abs}
\end{figure*}

\begin{figure*}
    \centering
    \includegraphics[width=0.69\columnwidth]{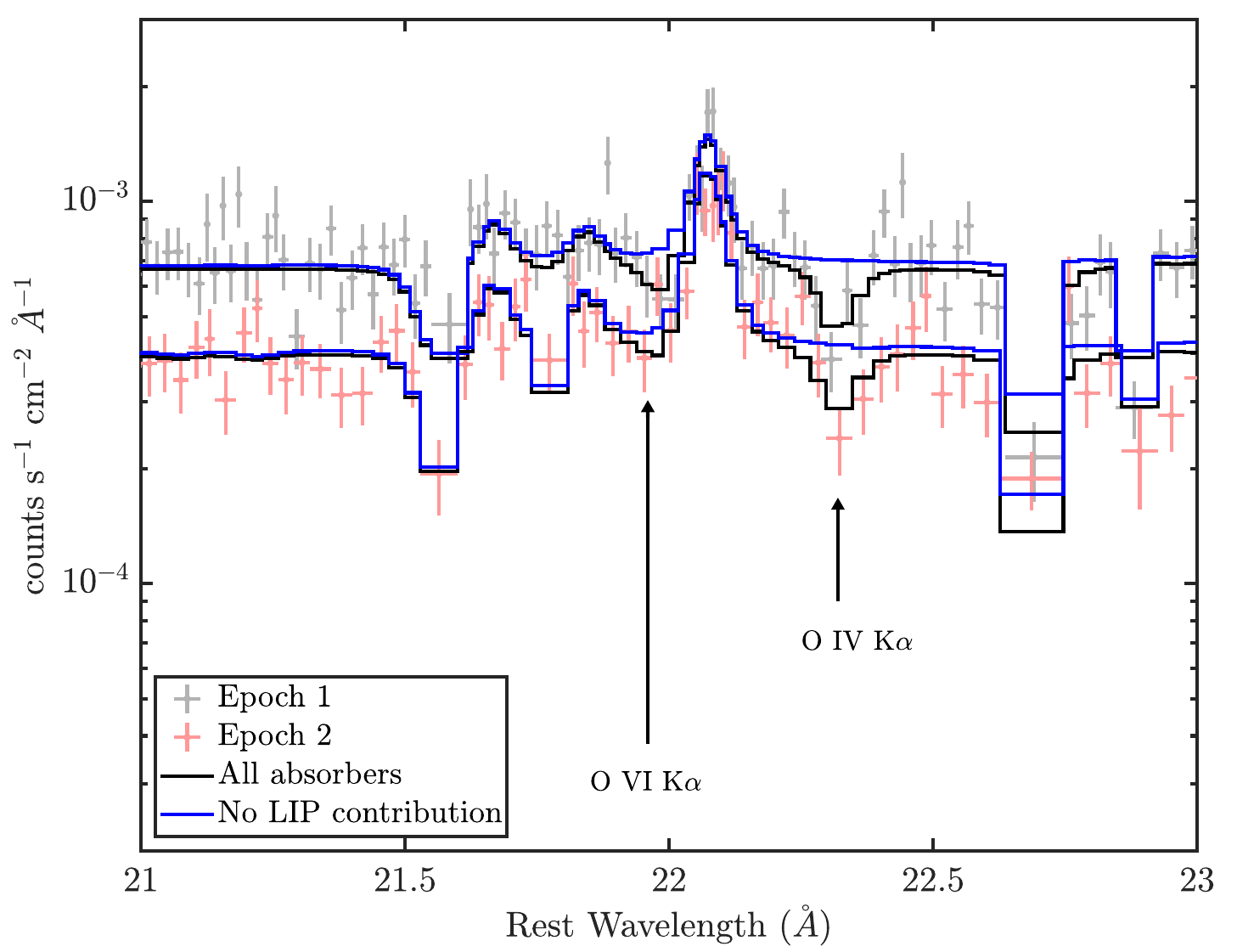}
    \includegraphics[width=0.69\columnwidth]{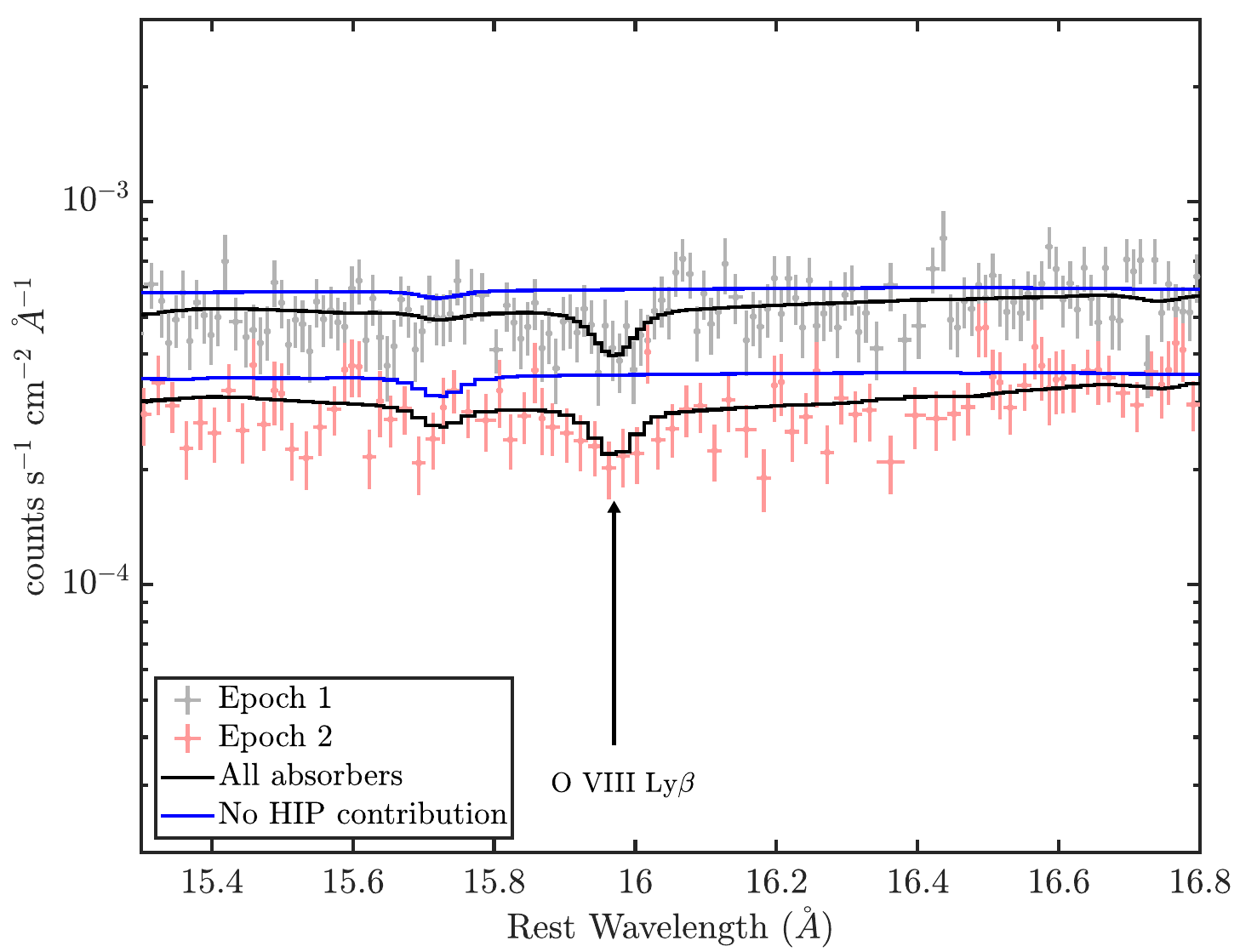}
    \includegraphics[width=0.69\columnwidth]{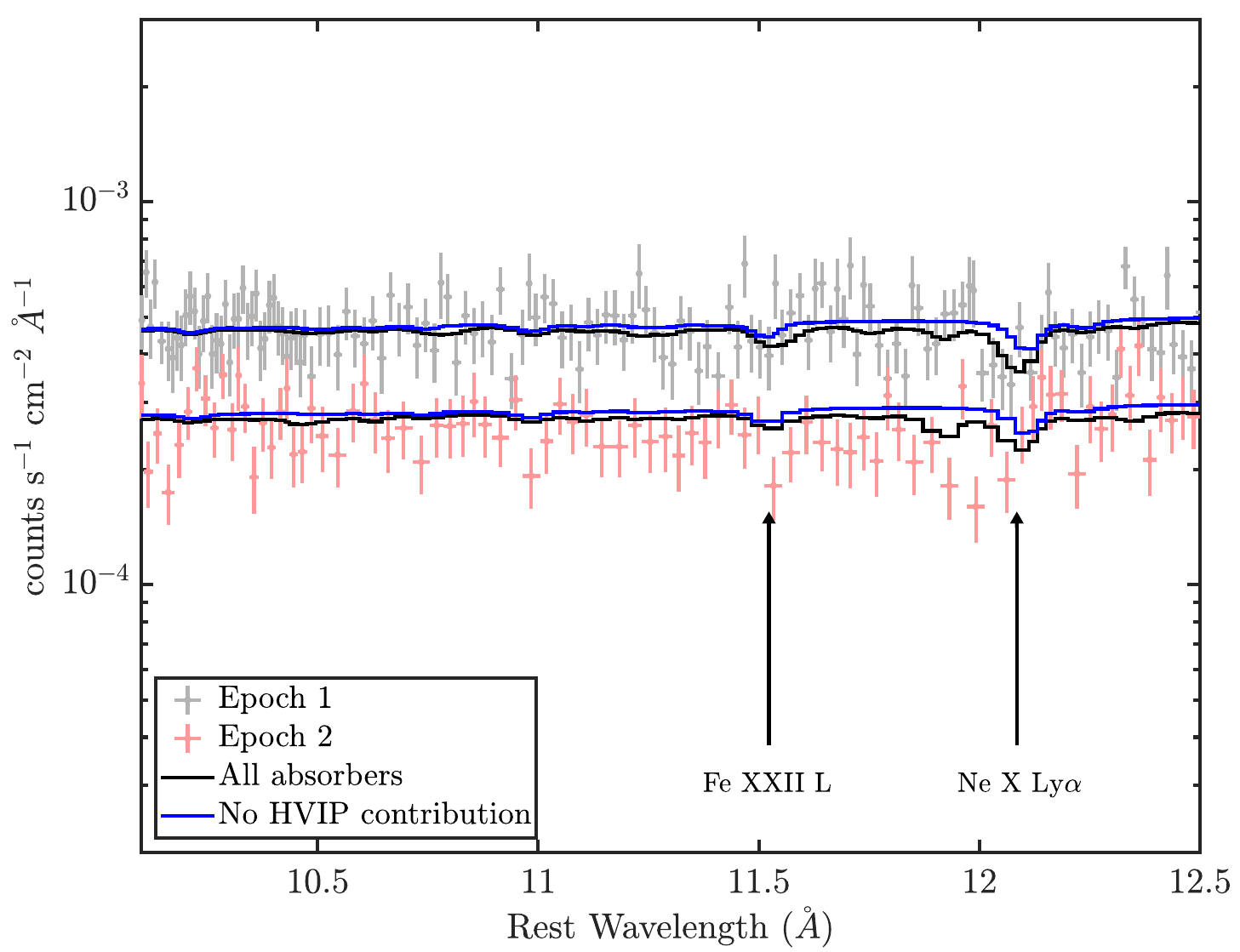}
    \caption{Examples of absorption lines for the contribution of the three absorbers. The grey points are the Epoch 1 observation, while the light red points are the Epoch 2 one. The black line represents the best-fit model listed in Table~\ref{tab:rgs_2abs}, while the blue one represents the model without one of the three absorbers, without refitting the model. Left panel. Spectral region in $\lambda=21-23$ \AA, the blue line was drawn by removing the LIP to highlight some of its contribution, namely O {\footnotesize IV} K$\alpha$ at $\lambda=22.76$ \AA\ and O {\footnotesize VI} K$\alpha$ at $\lambda=22.03$ \AA. Middle panel. Spectral region in $\lambda=15.3-16.8$ \AA, the blue line is obtained by removing the HIP, highlighting the strong contribution for O VIII Ly$\beta$ at $\lambda=16.01$ \AA. Right panel. Spectral region in $\lambda=10-12.5$ \AA, the blue line is obtained by subtracting the HVIP component, contributing for the Ne X Ly$\alpha$ at $\lambda=12.14$ \AA\ and for the Fe XXII L-transition at $\lambda=11.57$ \AA.}
    \label{fig:indilines}
\end{figure*}

The high-resolution spectrum of NGC 4051 obtained with the \textit{XMM-Newton} Reflection Grating Spectrometer (RGS) reveals the presence of three distinct ionized absorbers, each contributing uniquely to the observed absorption features. These absorbers, distinguished by their ionization levels and outflow velocities, imprint characteristic spectral signatures across the soft X-ray band.\\
\indent The LIP is characterized by a low ionization $\log U = -1.0\pm0.3$ and $\log U = -1.1^{+0.6}_{-0.4}$ for Epochs 1 and 2, respectively, corresponding to a photoionization-equilibrium temperature $T(LIP) \simeq 19,000$ K \citep[e.g.,][]{nicastro99,luminari23}, and a relatively low column density, of the order of $N_{\rm H} = 1.6^{+0.5}_{-0.6}\times 10^{20}$ cm$^{-2}$. This component imprints primarily K-shell absorption lines of moderately-ionized O {\footnotesize IV-VII}, observable at wavelengths $\lambda\sim21.5-22.5$ \AA~(see Fig.~\ref{fig:contrib_abs}), as well as weak inner L-shell unresolved lines of M-shell ions (Fe {\footnotesize XII-XVI}) at $\lambda\sim14-17$ \AA\, commonly referred to as Fe Unresolved Transition Arrays \citep[UTA, see e.g.,][]{behar01,krongold03}). \\
\indent The high-ionization phase (HIP) is the dominant contributor to the absorption structure. With $\log U = 1.23^{+0.07}_{-0.05}$ and $\log U = 1.12\pm0.04$ for Epochs 1 and 2, respectively, corresponding to a photoionization-equilibrium temperature $T(HIP)\simeq 1.5 \times 10^5$ K, and a higher column density of $N_{\rm H} = 1.4^{+0.2}_{-0.3}\times 10^{21}$ cm$^{-2}$, this absorber significantly affects the outer and inner L-shell transition of Fe {\footnotesize XVII-XXIV} and {\footnotesize XII-XVI} at $\lambda\sim7-15$ \AA\ and $\lambda\sim14-17$ \AA, respectively (see Fig.~\ref{fig:contrib_abs}). The outflow velocity of this phase ($v_{\rm out, HIP} = 530\pm70$ km s$^{-1}$) is consistent with the one of the LIP ($v_{\rm out,LIP} = 400\pm120$ km s$^{-1}$), which may imply that the two phases are co-spatial, likely in pressure equilibrium. If so, the LIP could be a cool condensed phase of the HIP, with number density  $n(LIP) = n(HIP) \times [T(HIP)/T(LIP)] \simeq 7.9\ n(HIP)$. Alternatively, the consistent velocity could be a coincidence and the LIP could be located farther away or closer to the black hole, depending on the density.\\
\indent Finally, the high-velocity and ionization phase (HVIP) is certainly kinematically distinguished from the LIP and HIP components, and represents the most extreme of the three absorbers, in terms of both ionization parameter ($\log U = 2.3^{+0.1}_{-0.2}$ for Epoch 1 and $\log U = 2.1^{+0.2}_{-0.3}$ for Epoch 2, corresponding to a photoionization-equilibrium temperature $T(HVIP)\simeq 8.5\times 10^5$ K) and column density $N_{\rm H} = 2.7^{+1.4}_{-1.6} \times 10^{21}$ cm$^{-2}$. The HVIP primarily imprints highly-ionized Fe L opacity and contributes to the line-like absorption at $\lambda\sim18.6$ \AA{} through O {\footnotesize VIII} K$\alpha$, where also the LIP O {\footnotesize VII} K$\beta$ plays a mild role (see Fig.~\ref{fig:contrib_abs}). Some individual lines for each absorber are shown in the three panels of Fig.~\ref{fig:indilines}.\\ 
\indent Overall, the presence of these three distinct absorbing phases highlights the complexity of the ionized gaseous environment of NGC 4051. Locating them will offer valuable insights into their relation with other ubiquitous components of AGN environments like broad and narrow emission line regions, molecular torus, etc., and their acceleration mechanisms.

\subsection{Are the absorbers in photoionization equilibrium?}
\begin{figure}
    \centering
    \includegraphics[width=\columnwidth]{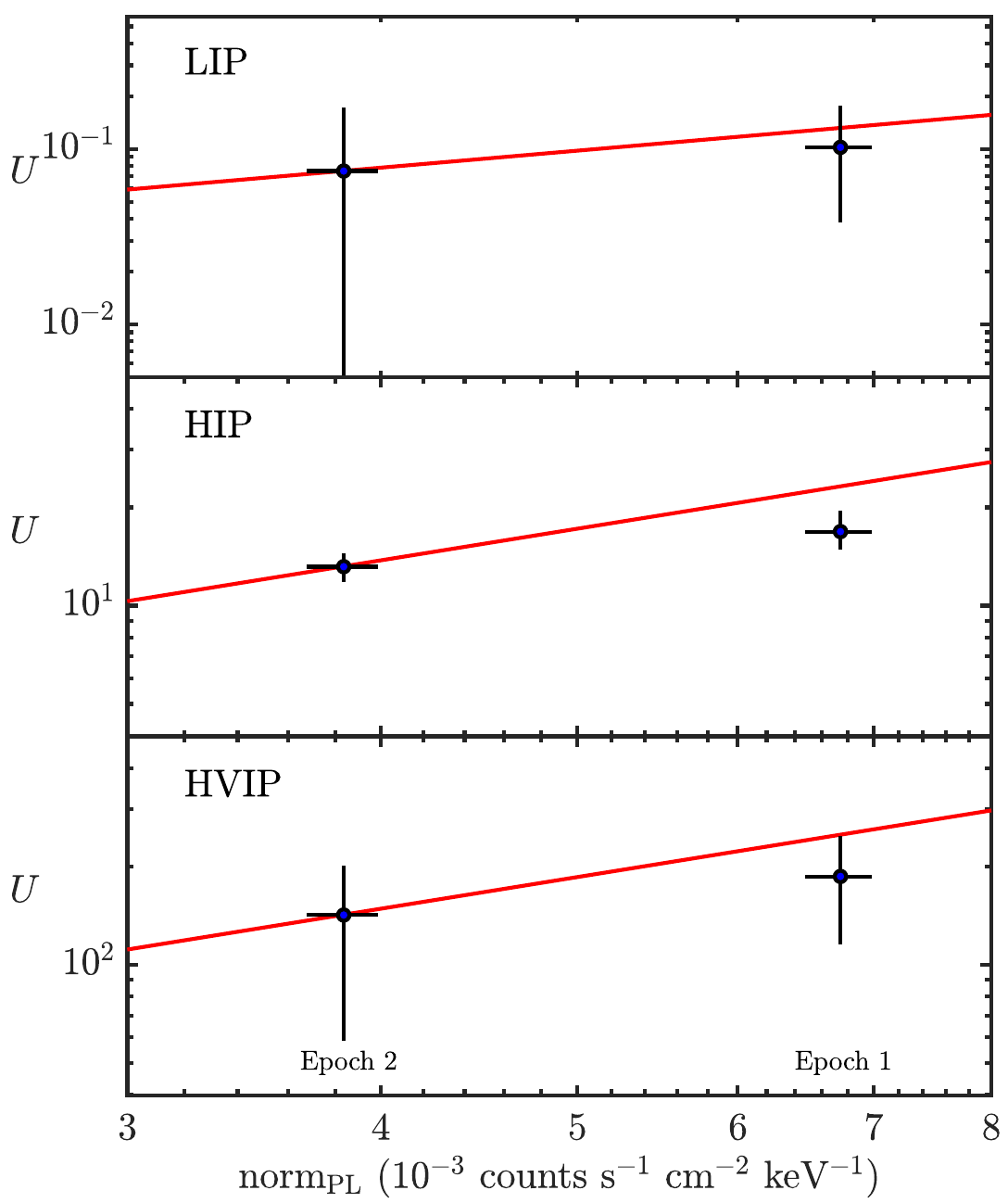}
    \caption{Ionization versus power law normalization of the three absorbers identified in this paper, each data point represents one Epoch. The red line is obtained assuming that the lowest flux state is in equilibrium.}
    \label{fig:equiplot}
\end{figure}



\indent As discussed in detail by \cite{nicastro99} and K07, the ionization state of a gas cloud in (instantaneous) photoionization equilibrium is uniquely determined, at any epoch, by the ionization parameter $U$ and, thus, must vary linearly with the rate of ionizing photons $Q_{\rm ion}$. Any deviation from this relationship would instead indicate a non-equilibrium state of the gas.\\ 
\indent In Fig.~\ref{fig:equiplot} we test the equilibrium hypothesis for the LIP, HIP and HVIP of NGC~4051, between the two {\it XMM-Newton} observing epochs, by plotting the best-fitting values of $\log U$ of these three components, as a function of the best-fitting X-ray power law normalization. Both the LIP and HVIP phases are consistent with being in photoionization equilibrium, though their ionization states also appear consistent with no variation at all, given the large uncertainties on their ionization parameter estimates. Normalizing the $U$ vs $Q_{\rm ion}$ relationship at Epoch~2 (with lower $Q_{\rm ion}$) yields values of $U$ during Epoch 1 (red line in Fig.~\ref{fig:equiplot}) consistent with those measured, within their $90\%$ confidence level. In contrast, the ionization state of the HIP significantly deviates from equilibrium conditions in Epoch~1 at a $3.8\sigma$ confidence level, suggesting a non-equilibrium ionization state.\\ 
\indent The linear relations in Fig.~\ref{fig:equiplot} (red lines), define estimates of the quantity $R^2n_{\rm H}$, i.e. the product between the gas density and the square of its distance from the ionizing source. These estimates are strictly valid only if the gas ionization state is consistent with the photoionization equilibrium hypothesis (i.e. LIP and HVIP), while represents only average estimates if the gas is not in instantaneous equilibrium with the ionizing source (i.e. the HIP). For the LIP, HIP and HVIP we obtain $R_{LIP}^2n_{\rm H}^{LIP}=4^{+4}_{-2}\times10^{40}$ cm$^{-1}$, $R_{HIP}^2n_{\rm H}^{HIP}=(1.7\pm0.2)\times10^{38}$ cm$^{-1}$, and $R_{HVIP}^2n_{\rm H}^{HVIP}=1.8^{+1.8}_{-0.7}\times10^{37}$ cm$^{-1}$, respectively.

\subsection{Time-evolving fit with TEPID}
\label{sec:tepid_fit}

\begin{figure}
    \centering
    \includegraphics[width=\columnwidth]{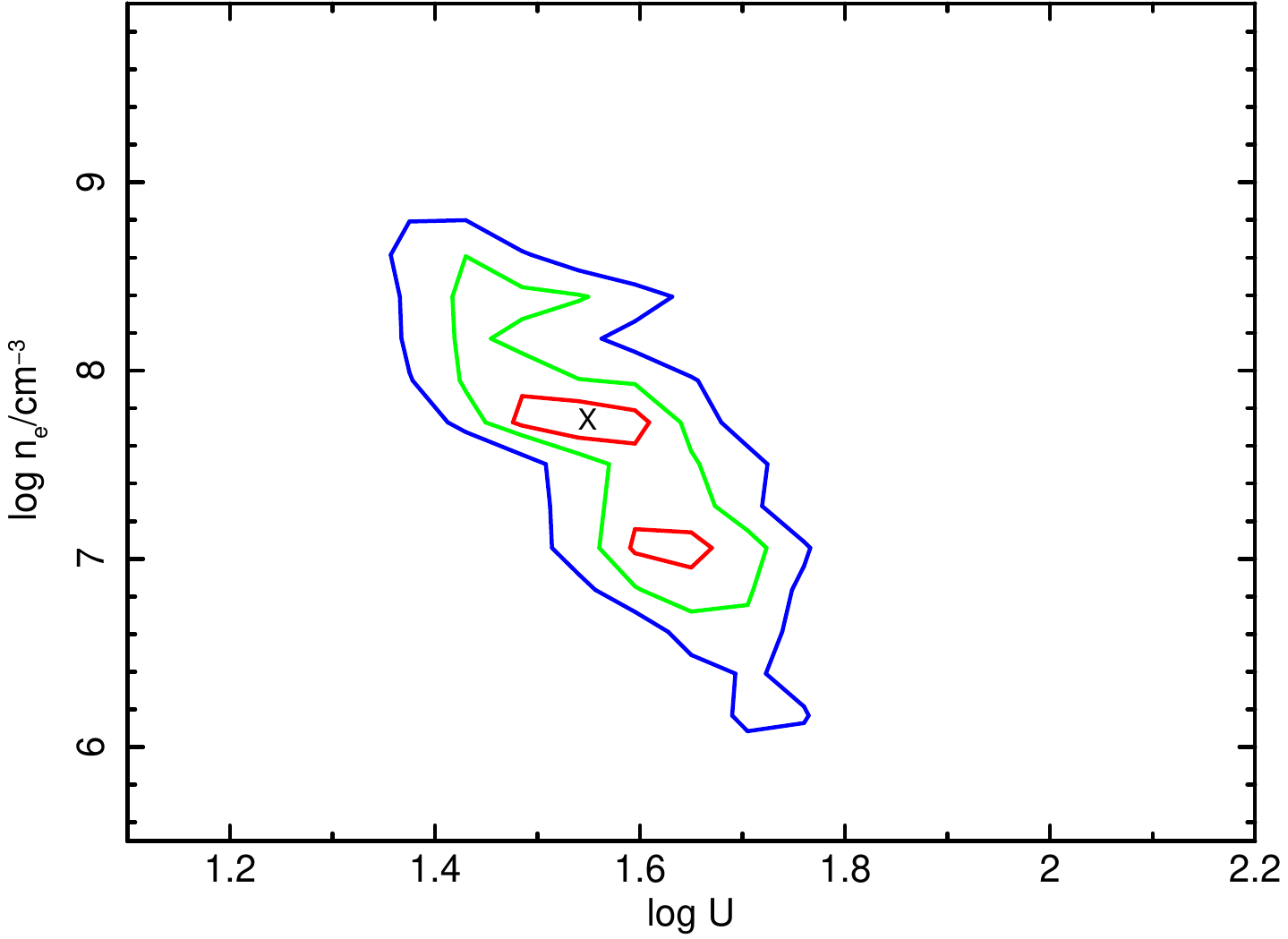}
    \caption{Two-dimensional confidence contours for the hydrogen number density $\log(n_{\rm H}/{\rm cm}^{-3})$ versus the ionization parameter $\log U$ for the HIP absorber, obtained from the {\footnotesize TEPID} fit. Contours correspond to the 68\% ($1\sigma$), 95.5\% ($2\sigma$), and 99.7\% ($3\sigma$) confidence levels for two parameters of interest. The black X marks the best-fit value.}
    \label{fig:tepidcont}
\end{figure}

The estimates of $R^2n_{\rm H}$ obtained under the photoionization-equilibrium hypothesis leave an intrinsic degeneracy between the gas number density and its distance from the ionizing source. This can only be removed by measuring independently one of these two quantities. One way to do this is using time-evolving photoionization models.\\
\indent In the extreme case in which the ionizing source variability timescale is much lower than the gas equilibration timescale, $t_{\rm var} \ll t_{\rm eq}$, implying low gas volume densities $n_{\rm H}$, the gas will never reach instantaneous equilibrium with the ionizing flux. Its ionization state will stay constant independently on the particular flux state, reflecting a sort of average photoionization-equilibrium state,  corresponding to the average flux of the ionizing source over the time. In this case, only an upper limit on $n_{\rm H}$ can be derived. Contrarily, when $t_{\rm var}\gg t_{\rm eq}$ (high $n_{\rm H}$), the gas will be in the photoionization equilibrium limit with the ionizing source at all epochs and only a lower limit on $n_{\rm H}$ can be derived. The most interesting case is when $t_{\rm var}\sim t_{\rm eq}$. In this case, the gas ionization state will try to follow and adapt to changes of the ionizing flux, but its response will be more or less smoothed and delayed, depending on the sign (increasing or decreasing flux) and amplitude of variability.\\ 
\indent Figure~\ref{fig:equiplot} shows that the ionization parameters of the LIP and HVIP, given their large uncertainties, are consistent with either remaining constant throughout the observing campaign or being in photoionization equilibrium between Epochs 1 and 2. This limits the possibility of constraining the density of these two components through time-evolving photoionization analysis. In contrast, the HIP shows a clear deviation from equilibrium. This allows us to remove the $N_{\rm H} - R$ degeneracy through time-evolving photoionization models.\\
\indent Here we use our Time-Evolving PhotoIonisation Device \citep[{\footnotesize TEPID},][]{luminari23}, which follows the gas ionization in time as a function of the time-variabile ionizing light curve. The resulting time-dependent ionic abundances are then interfaced to a custom version of {\footnotesize PHASE}, thus allowing one to directly fit time-resolved spectra. \\
\indent As every temporal integrator, TEPID needs the initial values at $t=0$ for all the variables to be integrated (i.e., the gas ionic distribution and temperature). We hereby assume that the gas cloud is in photoionization equilibrium at our $t=0$, the beginning of the {\it NuSTAR} observation, with an initial ionization parameter $U_0$\footnote{This hypothesis can be relaxed by allowing the additional parameter $P$ of the model that measures the $t=0$ over/under--ionization of the gas, compared to equilibrium, to be higher/lower than unity, respectively.}. We perform {\footnotesize TEPID} simulations covering the entire {\it NuSTAR} observation using the lightcurve from Fig. \ref{fig:lc}. We adopt the standard {\footnotesize CLOUDY} AGN template for the SED from $10^{-3}$ eV (the lower bound of TEPID) up to the OM point at $7$ eV. Between $0.3$ and $100$ keV (the upper bound of TEPID), we use the continuum best-fit of the {\it XMM-Newton} EPIC-pn and {\it NuSTAR} spectra, rescaled to the {\it NuSTAR} flux at $t=0$. The two SED segments are connected with a power law to ensure continuity. At each time step, the low-energy part of the SED is kept fixed, while the $0.3$–$100$ keV segment of each observation is rigidly scaled according to the light curve, with the connecting power law recomputed accordingly. We span the values of $\log(U_0)$ from $-2$ to $2$ with steps of 0.125 and $\log(n_{\rm H}/cm^{-3})$ from 5 to 10 with 0.5 steps. Simulations are run from $\log(N_{\rm H}/cm^{-2})=20$ (as the gas is optically thin below this value) to 22 with 0.05 steps. Further details will be provided in Paper II.\\
\indent We then compute the flux-weighted average ionic columns throughout the {\it XMM-Newton} Epochs 1 and 2 and we input them to our custom spectral interface. This way, we replace the LIP, HIP, HVIP equilibrium components with three {\footnotesize TEPID} components and refit the data.
The resulting best-fitting model yields a fit statistic of $C/{\rm dof} = 6622/6258$. The LIP and HVIP best-fitting $\log U_0$ are $\log U_0 = -0.3_{-0.3}^{+0.2}$ and $\log U_0>2.05$, respectively, both consistent, within their large $95\%$ uncertainties, with the best-fitting values measured under the photoionization equilibrium assumption during the two epochs and listed in Table~\ref{tab:rgs_2abs}. As anticipated, $n_{\rm H}$ cannot be constrained for LIP and HVIP, neither upper or lower limits are found for those absorbers. For the HIP component we get $\log U_0 = 1.55^{+0.08}_{-0.10}$ and $\log(n_{\rm H}/{\rm cm}^{-3})=7.7^{+0.2}_{-0.9}$, both constrained at $3\sigma$ confidence level (Fig.~\ref{fig:tepidcont}).\\
\indent Using this updated value of $U_0$ we now obtain $R^2 n_{\rm H} = 6^{+2}_{-1} \times 10^{37}~{\rm cm}^{-1}$, a factor of $\sim3$ smaller than what estimated under the incorrect equilibrium assumption. We finally derive a distance of $R = 0.45^{+0.80}_{-0.09}$~light-days, corresponding to $3.7^{+6.7}_{-0.8} \times 10^{-4}$~pc$=4000^{+7000}_{-800}R_{\rm g}$, marginally consistent with the location of the high-ionization broad emission line clouds \citep[][]{peterson00}. This result is remarkably consistent with the measurement of K07, who found $R\sim0.5\div1.0$ light days. Remarkably, this distance is several orders of magnitude smaller than the minimum distance inferred assuming that the measured velocity is the escape velocity \citep[e.g.,][]{crenshaw12,serafinelli19}. For $v_{\rm out}\simeq500$ km/s we find $R\simeq4\times10^5\;R_g$, which is around 100 times larger than the value we found. This means that the observed gas is either destined to fall back down towards the inner region, as a failed wind, or, alternatively, our line of sight is probing a portion of the wind (e.g. its basis) that has yet to be fully accelerated outwards \citep[e.g.,][]{elvis00}. Our result also aligns well with the time-evolving photoionization analysis presented by \citet{sadaula25}, who used a different modeling approach with a different data set and found $\log(n_{\rm H}/{\rm cm}^{-3}) > 6.95$ and $R < 0.02$~pc. However, they only considered two ionized absorbers. One of the two components is characterized by $\log\xi/({\rm erg\;cm\;s^{-1}})\simeq1.9$ ($\log U\sim0.4$) and by a velocity $v_{\rm out}\sim3500$ km s$^{-1}$. Given the low ionization, intermediate between that of the LIP and HIP found in this paper, it is likely a superposition of the two low-velocity absorbers, with the higher velocity arising from the lower energy resolution of NICER with respect to RGS. The other component is characterized by $\log\xi/({\rm erg\;cm\;s^{-1}})\simeq3.4$ ($\log U\simeq1.9$) and $v_{\rm out}\simeq10,000$ km s$^{-1}$. The ionization is remarkably similar to the HVIP component in our paper (see Section~\ref{sec:hvip_comparison}), which is supported by the large detected velocity. These converging results strengthen the interpretation of the HIP absorber as a compact, dense structure located close to the ionizing source.

\subsection{Origin and Stability of Soft X-ray Emission Lines}
\label{sec:emission_lines}

The soft X-ray spectrum of NGC 4051 is characterized by prominent emission lines from O {\footnotesize VII}, O {\footnotesize VIII}, and Ne {\footnotesize IX}, which remain constant over the $\sim250$ ks timescale probed by our observations, from the beginning of Epoch 1 to the end of Epoch 2. This is fully consistent with previous findings from  \cite{steenbrugge09}, who reported no short-term variability of the soft X-ray emission lines of NGC 4051 during a 250 ks long {\em Chandra}-HETG observation of the source.\\ 
\indent This lack of variability suggests that these emission lines originate in a region that is not directly influenced by the short-term changes in the ionizing continuum ($d>3$ light days), and rule out both the inner accretion disk or the more variable warm absorbers, as responsible for the emissivity. 
This is also consistent with numerous previous studies that associate the soft X-ray emission in AGN with the narrow-line region (NLR), which extends over tens to hundreds of parsecs from the central black hole and is composed of low-density, photoionized gas \citep[e.g.,][]{kinkhabwala02,bianchi06, guainazzi07, bianchi10, grafton-waters21,braito17}.\\
\indent A previous study on this source reported the presence of a broad component in the soft X-ray emission lines, with a turbulent velocity of $\sim4000$ km s$^{-1}$ \citep{peretz19}. This feature was interpreted as emission from the broad-line region (BLR) of NGC~4051. To investigate the presence of a similar component in our data, we added a second Gaussian profile to each of the five emission lines, fixing the width to $v_{\rm turb} = 3800$ km s$^{-1}$ and tying the centroid to that of the corresponding narrow component. Line normalizations were left free to vary. After the addition of the broad components, the total flux of each line is approximately evenly split between the narrow and broad Gaussians. However, the improvement in the fit is modest ($\Delta C/\Delta{\rm dof} = 20/9$), suggesting that the broad components might be present, but they are not detected at high significance. Furthermore, the normalization of the broad Gaussians remains consistent between Epochs 1 and 2 at the 90\% confidence level. We note that the observations analyzed by \citet{peretz19} correspond to a significantly lower flux state, with a power-law normalization $\sim$2–3 times lower than in our Epochs~2 and~1, respectively. This difference may affect the detectability of BLR emission, as its contrast with respect to the continuum would be higher during a low-flux state.

\subsection{Comparison with Previous Studies on High-Velocity Ionized Absorber}
\label{sec:hvip_comparison}

\begin{table}
\centering
\caption{Comparison of HVIP absorber parameters in NGC 4051 with previous observations in literature. We converted $\log U$ into $\log \xi$ using the correction factor proposed by \cite{crenshaw12}, $\log U=\log\xi-1.5$, which however is only valid under the assumption that $\xi$ and $U$ were computed using the same SED.}
\label{tab:hvip_comparison}
\begin{tabular}{lccc}
\hline \hline
Reference & $v_{\rm out}$ (km s$^{-1}$) & $\log \xi$ (erg cm s$^{-1}$)\\
\hline
\cite{steenbrugge09} & $4670\pm150$ & $3.1\pm0.1$\\
\cite{lobban11} & $5800^{+860}_{-1200}$ & $4.1^{+0.2}_{-0.1}$\\
\cite{sadaula25} & $10900\pm1200$ & $3.44\pm0.05$\\
This work (Epoch 1) & $5800\pm300$ & $3.8^{+0.1}_{-0.2}$\\
This work (Epoch 2) & $-$ & $3.6^{+0.2}_{-0.3}$\\
\hline
\end{tabular}
\end{table}

The detection of the high-velocity and ionization phase (HVIP) in the present analysis, which was undetected in K07, is consistent with earlier studies reporting fast, highly ionized outflows in NGC~4051.\\
\indent \cite{lobban11} analyzed a combined 300 ks observation using the \textit{Chandra} High Energy Transmission Grating (HETG) and the \textit{Suzaku} X-ray Imaging Spectrometer (XIS), and reported a highly ionized outflow component characterized primarily by Fe {\footnotesize XXV} and Fe {\footnotesize XXVI} K-shell absorption lines at observed energies around $\sim 6.8$ and $\sim 7.1$ keV. They measured an outflow velocity of $v_{\rm out}=5800^{+860}_{-1200}$ km s$^{-1}$, along with a column density of $N_{\rm H} = 8.4^{+1.9}_{-2.0}\times10^{22}$ cm$^{-2}$, and an ionization parameter of $\log \xi = 4.1^{+0.2}_{-0.1}$. Adopting the $U-\xi$ conversion from \cite{crenshaw12}, $\log U \simeq \log \xi - 1.5$, the corresponding ionization parameter from \citet{lobban11} would be $\log U \simeq 2.6^{+0.2}_{-0.1}$. While this value is slightly higher than the one we measure ($\log U=2.1-2.3$ in Epochs 1 and 2, respectively), discrepancies can naturally arise due to differences in the assumed spectral energy distributions (SEDs) and modeling approaches between the studies.\\
\indent Similarly, \citet{steenbrugge09}, analyzing a 250 ks \textit{Chandra} Low Energy Transmission Grating Spectrometer (LETGS) observation, reported a high-velocity absorber ($v_{\rm out}=4670\pm150$ km s$^{-1}$) with an ionization parameter $\log \xi=3.19\pm0.09$ in units of erg s$^{-1}$ cm and a column density of $N_{\rm H}=(2\pm1)\times10^{22}$ cm$^{-2}$. Thus, applying the same log~$U$-log~$\xi$ conversion, the corresponding ionization parameter from \citet{steenbrugge09} is $\log U=1.7\pm0.1$.\\
\indent Finally, \cite{sadaula25} found an absorbing component characterized by $\log\xi/({\rm erg\;cm\;s^{-1}})\simeq3.4$ ($\log U\simeq1.9$ with the usual conversion) and $v_{\rm out}\simeq10,000$ km s$^{-1}$, with a column density of $N_{\rm H}\simeq10^{22}$ cm$^{-2}$.
The ionization state of our HVIP absorber ($\log U=2.1-2.3$) lies therefore in between those of \citet{lobban11}, \citet{sadaula25} and \citet{steenbrugge09}, reinforcing the interpretation of the HVIP outflow as a persistent component of the NGC~4051 outflow system. Variations observed in the column density of this component may reflect the inherent complexity and multi-phase nature of the absorber, possibly indicating density stratification. For clarity, we summarize these comparisons in Table~\ref{tab:hvip_comparison}.\\ 
\indent The detection of the HVIP component in this work, along with previous observations by \citet{steenbrugge09} and \citet{lobban11}, suggests that this fast and highly ionized outflow is a persistent feature of NGC 4051 over timescales of at least a decade. Assuming a constant outflow velocity of $v_{\rm out}\sim5800$ km s$^{-1}$, the HVIP would have traversed a distance of $d\sim0.05$ pc in 10 years, corresponding to $\sim1.5\times10^{17}$ cm. This opens up several intriguing physical scenarios regarding its nature and location relative to the HIP component.\\
\indent One possibility is that the HVIP originates at smaller radii than the HIP and propagates radially outward, eventually shocking against a denser external medium, possibly the BLR or a failed wind. In this framework, the HIP could represent the cooling, photoionized aftermath of such a shock. While speculative, this picture resonates with the shocked wind feedback models invoked in the context of quasar outflows \citep[e.g.,][]{zubovas12,king15}.\\
\indent Alternatively, if the outflows are not radial but instead have a significant transverse velocity component, perhaps due to a wide-angle wind emerging from the accretion disk as proposed in K07, then the HVIP, HIP, and LIP may all represent coexisting layers of a stratified, multi-phase wind seen at different launching radii. In this case, the faster HVIP would naturally arise from smaller disk radii, while the HIP and LIP, which share consistent velocities over more than a decade, may be associated with longer-lived transverse flows. This would also explain the relatively stable ionization and velocity properties of the HIP and LIP across multiple observations.\\
\indent A third, less likely possibility is that the HVIP lies at greater distances than the HIP. However, the higher velocity and ionization level of the HVIP make this configuration counterintuitive. Furthermore, if the HVIP were a freely expanding, non-replenished shell of gas moving at terminal velocity, its column density would be expected to decline with time due to geometric dilution. This might explain the difference in column densities between our detection and that of \citet{steenbrugge09} and \citet{lobban11}, assuming the ionization state remains stable due to decreasing density compensating for declining ionizing flux.

\section{Conclusions}
\label{sec:conclusions}

\noindent We have conducted a high-resolution spectral analysis of two \textit{XMM-Newton} observations of NGC~4051, identifying and characterizing three distinct ionized absorbers. We have constrained their physical properties, studied their photoionization equilibrium, and tested time-dependent models with {\footnotesize TEPID}. We summarize our main findings below.\\
\begin{itemize}
\item We detect three ionized absorbers in the high-resolution X-ray spectrum of NGC 4051: a low-ionization phase (LIP), a high-ionization phase (HIP), and a high-velocity, high-ionization phase (HVIP), each with distinct ionization levels, column densities, and outflow velocities.
\item The three phases contribute differently to the RGS spectrum: the LIP produces O\,\textsc{iv} -\,\textsc{vii} and some Fe L-shell UTA absorption in Fe M-shell ions; the HIP dominates the Fe L inner- and outer-shell regions ($7-17$\,\AA); the HVIP accounts for residual opacity in the outer-shell Fe L region and a fast O\,\textsc{viii} K$\alpha$ line.
\item Under the ionization equilibrium assumption, the product $R^2 n_{\rm H}$ is determined for each absorber. While the LIP and HVIP are consistent with equilibrium across both epochs, the HIP exhibits significant deviations from equilibrium.
\item We apply the time-dependent photoionization code {\footnotesize TEPID}, to constrain the number density of the HIP to $\log(n_{\rm H}/{\rm cm}^{-3}) = 7.7^{+0.2}_{-0.9}$, leading to a distance of $R = 0.45^{+0.80}_{-0.09}$ light-days, or $R=3.7^{+6.7}_{-0.8}\times10^{-4}$~pc$=4000^{+7000}_{-800}\;R_{g}$.
\item Our best-fitting HVIP parameters agree well with those of a fast outflow component reported in previous studies, including \citet{steenbrugge09} and \citet{lobban11}, confirming the persistence of this absorber over more than a decade. This long-term stability supports a multi-phase, stratified wind scenario. The detection of the HVIP in the soft X-ray band, in contrast to earlier Fe K-band detections, reveals that it imprints features across a wide energy range and may be part of a transverse outflow originating from the inner accretion disk.
\item The intensity of the soft X-ray emission lines from O\,\textsc{vii}, O\,\textsc{viii}, and Ne\,\textsc{ix} remains  constant over the 250 ks campaign, consistent with an origin in the narrow-line region and not associated with the rapidly varying inner absorber zones. Only marginal evidence of broad lines, previously detected in \cite{peretz19}, was found in these observations.
\end{itemize}

These results highlight the power of high-resolution X-ray spectroscopy in unveiling the structure and variability of ionized outflows in nearby AGN. While our analysis confirms the presence of multiple absorbing phases in NGC~4051, the departure from equilibrium observed in the HIP phase suggests that time-dependent ionization modeling is essential to fully characterize the physical properties of AGN winds. Tools such as {\footnotesize TEPID} offer a promising method for deriving absorber densities and locations, even when only a few epochs are available.\\
The upcoming availability of new-generation X-ray observatories will further revolutionize this field. In particular, \textit{XRISM} will be able to provide high resolution soft X-ray spectra with unprecedented sensitivity, once the gate valve protecting Resolve is opened, enabling precise measurements of absorber line profiles and physical conditions. On longer timescales, the X-IFU instrument on board \textit{NewAthena} \citep{cruise25} will combine exquisite spectral resolution with large collecting area, enabling time-resolved spectroscopy of warm absorbers on intra-observation timescales. These capabilities will be critical to disentangle the physical mechanisms driving AGN outflows and their variability, and to map their impact on the surrounding environment with greater accuracy.\\
In Paper~II (Luminari et al., in prep.), we will extend this work by analyzing time-resolved spectra from the EPIC-pn instrument. Although EPIC-pn has a lower spectral resolution compared to the RGS, its significantly larger effective area enables high-quality spectroscopy of a much larger number of intervals. This will allow us to track the evolution of the ionization parameter on shorter timescales, providing a complementary view of the absorber variability and further constraining the physical response of the outflowing gas to changes in the ionizing continuum.\\

\acknowledgments The authors thank the referee for their valuable comments that improved the quality of this paper. RS acknowledges funding from the CAS-ANID grant number CAS220016. RS and FN acknowledge financial support from INAF-PRIN grant “A Systematic Study of the largest reservoir of baryons and metals in the Universe: the circumgalactic medium of galaxies” (No. 1.05.01.85.10). FN, FC and EP acknowledge financial support from PRIN-MUR-2022 grant “Advanced X-ray modeling of black hole winds” (DRAGON)” (No. PRIN 2022K9N5B4). YK acknowledges support from DGAPA-PAPIIT grant IN102023. LP, FN and AL acknowledge financial support from EU HORIZON-2020 grant “AHEAD2020” (Agreement No. 871158). LP and FN acknowledge support by ASI (Italian Space Agency) through the Contract no. 2019-27-HH.0 on Athena.\\
The research is partly based on observations obtained with XMM–Newton, an ESA science mission with instruments and contributions directly funded by ESA Member States and NASA. This research has made use of data and software provided by the High Energy Astrophysics Science Archive Research Center (HEASARC), which is a service of the Astrophysics Science Division at NASA/GSFC and the High Energy Astrophysics Division of the Smithsonian Astrophysical Observatory. This research has made use of the NuSTAR Data Analysis Software (NUSTARDAS), jointly developed by the ASI Space Science Data Center (SSDC, Italy) and the California Institute of Technology (Caltech, USA).

\bibliographystyle{likeapj}
\bibliography{biblio}    


\end{document}